\documentstyle[twocolumn,eqsecnum,aps,epsf,epsfig]{revtex}

\newcommand{\be}{\begin{equation}}
\newcommand{\ee}{\end{equation}}
\newcommand{\bea}{\begin{eqnarray}}
\newcommand{\eea}{\end{eqnarray}}

\newcommand{\lton}{\mathrel{\lower.9ex
                  \hbox{$\stackrel{\displaystyle <}{\sim}$}}}

\newcommand{\nn}{\nonumber}

\newcommand{\mat}{\left ( \begin{array}{cc}}
\newcommand{\emat}{\end{array} \right )}
\newcommand{\matt}{\left ( \begin{array}{ccc}}
\newcommand{\ematt}{\end{array} \right )}
\newcommand{\matf}{\left ( \begin{array}{cccc}}
\newcommand{\ematf}{\end{array} \right )}
\newcommand{\vect}{\left ( \begin{array}{c}}
\newcommand{\evect}{\end{array} \right )}

\def\beqn{\begin{eqnarray}}
\def\eeqn{\end{eqnarray}}

\def\d{\partial}

\begin{document}

\title{Dashen's Phenomenon in Gauge Theories with Spontaneously Broken 
Chiral Symmetries}
\author{G.\ Akemann$^{a}$
%{Unit\'e associ\'e CNRS/SPM/URA 2306}
, J.T.\ Lenaghan$^{b}$ and K.\ Splittorff$\,^b$}
\address{
$^a$ CEA/Saclay, Service de Physique 
Th\'eorique,
F-91191 Gif-sur-Yvette Cedex, France\\
$^b$ The Niels Bohr Institute, Blegdamsvej 17, 
DK-2100 Copenhagen {\O}, Denmark}
%\date{\today}
%\begin{flushleft}
%SPhT-T01/102
%\end{flushleft}

\maketitle   

%%%%%%%%%%%%%%%%%%%%%%%%%%%%%%%%%%%%%%%%%%%%%%%%%%%%%%%%%%%%%%%%%%%%%%%%

\begin{abstract}
We examine Dashen's phenomenon in the
Leutwyler--Smilga regime of QCD with any number of colors and quarks
in either the fundamental or adjoint representations of the gauge
group. In this limit, the theories only depend on simple combinations
of quark masses, volume, chiral condensate and vacuum angle. 
Based upon this observation, we derive simple expressions for   
the chiral condensate and the topological density 
and show that they are in fact related. By examining the
zeros of the various partition functions, we elucidate the mechanism
leading to Dashen's phenomena in QCD.
\end{abstract}
%\pacs{PACS numbers: 11.30.Rd, 11.30.Qc, 12.39.Fe}

\narrowtext      

%%%%%%%%%%%%%%%%%%%%%%%%%%%%%%%%%%%%%%%%%%%%%%%%%%%%%%%%%%%%%%%%%%%%%%%

\section{Introduction}

The vacuum angle, $\theta$, in QCD is experimentally constrained to be
zero with a deviation of less than $10^{-9}$
\cite{Smith:1990ke}.  This amounts to a theoretical
fine-tuning problem which is conceptually similar to the vanishing of
the cosmological constant.  Without an understanding of the physics at
nonzero values of $\theta$, one will probably not be able to explain
why $\theta$ is essentially zero.  In this work, we approach the
properties of the $\theta$-vacua from the vantage point of the hadronic
phase where the physics is determined by the spontaneous breaking of
chiral symmetry (SB$\chi$S).  We consider QCD in the Leutwyler--Smilga
regime \cite{Leutwyler:1992yt} 
for which the volume, $L^4$, of Euclidean space-time is taken such that
\be
\frac{1}{\Lambda_{{\rm QCD}}} \ll L \ll \frac{1}{m_\pi}.
\ee
The lower limit ensures that chiral perturbation theory is valid and
that the Goldstone modes associated with %the 
SB$\chi$S are the dominant degrees of freedom. The upper bound
ensures that the Compton wavelength of the Goldstone modes is much
greater than the size of the box and thus these modes can be treated as
constant \cite{Gasser:1987ah}.  
This has the advantage of allowing for exact, 
analytic calculations.% \cite{Leutwyler:1992yt}.

The various patterns of the SB$\chi$S depend on both the number of
colors, $N_c$, and the representation of the 
$N_f$ matter fields. They are simply
codified by the Dyson index, $\beta$, which is assigned according to
the anti-unitary symmetry of the Euclidean Dirac operator
\cite{Verbaarschot:1994qf}.  For $N_c \ge 3$ and matter in the
fundamental representation ($\beta=2$), the pattern of the SB$\chi$S
is given by $SU(N_f)\times SU(N_f)\rightarrow SU(N_f)$.  Matter in the
fundamental representation with $N_c=2$ ($\beta=1$) has the 
pattern $SU(2 N_f)\rightarrow Sp(2 N_f)$.  Finally, for any $N_c$
with matter in the adjoint representation ($\beta=4$), the 
symmetry breaking pattern is given by $SU(N_f)\rightarrow SO(N_f)$.

At the level of the chiral QCD Lagrangian, the theory depends only on
the combination $M e^{i\theta/N_f}$ where $M$ is the quark mass
matrix \cite{DiVecchia:1980,Witten}. 
In the Leutwyler--Smilga regime \cite{Leutwyler:1992yt}, this
dependence is constrained even further.  The quark masses, $m_j$, occur
only in the rescaled combination $\mu_j=m_jV\Sigma$, where $V$ is the
volume and $\Sigma$ is the chiral condensate in the chiral limit which
is assumed to be non-vanishing.  For one and two degenerate quark
flavors Leutwyler and Smilga and Verbaarschot
\cite{Leutwyler:1992yt,Smilga:1995tb} found by explicit calculation that the 
exact  partition function only depends on the combination $\mu
\cos(\theta/2)$.

One striking aspect of QCD at nonzero values of $\theta$ is the
spontaneous breaking of CP in a first-order phase transition at
$\theta=\pi$ known as Dashen's phenomenon \cite{Dashen}.  
In \cite{DiVecchia:1980,Witten}, this was shown 
by examining the $\theta$-dependence of the energy density of 
the large-$N_c$ chiral Lagrangian
and was reconsidered more recently in \cite{Creutz,Smilga,Tytgat}
for different numbers of flavors and mass splittings.
Going to the Leutwyler--Smilga regime, it was shown in Ref.\
\cite{Lenaghan:2001ur} for $\beta=2$ that 
Dashen's phenomenon is indeed realized 
by directly calculating the 
discontinuity in the topological density at $\theta=\pi$.  
In this work, we extend \cite{Lenaghan:2001ur} to $\beta=$ 1 and 4
by calculating the corresponding partition function, 
chiral condensate and topological density. 
We demonstrate that Dashen's phenomenon is generic to QCD
regardless of the pattern of SB$\chi$S.  By studying the zeros of the
partition functions, we show that this is a direct consequence of the
very restricted dependence of the partition function on quark masses,
the chiral condensate and $\theta$. This restriction also establishes
a direct relationship between the chiral condensate and the
topological density as well as between their 
susceptibilities. The $N_f=2$ theories are sufficiently
rich to illustrate the above points and as such we focus mainly on these
cases. For a discussion of $N_f>2$, see the appendix.

%%%%%%%%%%%%%%%%%%%%%%%%%%%%%%%%%%%%%%%%%%%%%%%%%%%%%%%%%%%%%%%%%%%%

\section{Partition functions}
\label{sec:Z1}

In order to study Dashen's phenomenon for generic patterns of the
SB$\chi$S, we first calculate the partition functions for $\beta=$1
and 4 including the contributions from all topological sectors.  The
full partition function as a function of $\theta$ may be decomposed
into a sum over partition functions each of which is 
restricted to a fixed topological charge, $\nu$,
\be
{\cal Z}^{(\beta)}(\theta,\{ \mu_i \}) = \sum_{\nu=-\infty}^{\infty} 
e^{i\nu\theta}{\cal Z}^{(\beta)}_\nu(\{ \mu_i \}) \ .
\ee
Expressions at fixed $\nu$
%${\cal Z}^{(2)}_\nu(\{ \mu_i \})$ 
were derived
%in general 
in Refs.\ \cite{Brower:1981vt} for $\beta=2$ 
and for degenerate masses with $\beta=1$ and 4 in Ref.\ \cite{Smilga:1995tb}.
Recently, 
the authors of Refs.\ \cite{Nagao:2000qn,Akemann:2000} were able to
calculate expressions for non-degenerate masses using 
chiral Random Matrix Theory ($\chi$RMT).
%${\cal Z}^{(\beta)}_\nu(\{ \mu \})$ with
%$\beta=1$ and $4$ (including certain mass-degeneracies) within the
%chiral random matrix theory.  
The equivalence between $\chi$RMT 
and QCD in the Leutwyler--Smilga regime has been
established in a number of works (see Ref.\ \cite{Verbaarschot:2000dy}
for a review of the literature).  Using the results of 
\cite{Nagao:2000qn,Akemann:2000}, we now
calculate the full partition functions including the contributions
from all topological sectors.  Our normalization will be such that at
vanishing masses
${\cal Z}^{(\beta)}(\theta,\{ \mu\!=\!0 \})=1$, 
which is only possible after summing over all topological charges \cite{Poul:1999}.
The results for $N_f=1$ are known and are independent of $\beta$ 
\cite{Leutwyler:1992yt,Smilga:1995tb}:
\be
{\cal Z}^{(\beta)}(\theta,\mu) = e^{\mu \cos\theta} \ .
\label{Nf=1}
\ee
For $\beta=4$, one must replace $\theta\to\theta/N_c$ . 
%the result is
%\be
%{\cal Z}^{(4)}(\theta,\mu)= e^{\mu \cos\theta/N_c} \ .
%\ee

%\subsection{${\fett \beta=1}$}

We begin with the first non-trivial case of two non-degenerate 
flavors. The same technique applies to $N_f>2$ where we refer to the appendix.
The $\beta=1$ partition function with fixed topological charge is 
\cite{Nagao:2000qn}
\bea \nonumber
{\cal Z}^{(1)}_\nu(\mu_1,\mu_2) &=& 8\int_0^1 \!dt\ t^2        
\Big\{ I_{\nu}(t\mu_1) \mu_2I_{\nu-1}(t\mu_2) \\ 
    &-&I_{\nu}(t\mu_2) \mu_1I_{\nu-1}(t\mu_1) \Big\} \ .
\label{Z1nu}
\eea
In order to sum over all topological charges we use the identity
\bea
&&\sum_{\nu=-\infty}^\infty e^{i\nu\theta} I_{\nu+n}(\mu_1)I_{\nu+m}(\mu_2) =
\label{master}\\
&&=e^{-in\theta}
\left(\frac{\mu_{12}(\theta)}{\mu_1e^{-i\theta}+\mu_2}\right)^{n-m} 
%\nonumber\\
%&\times&
I_{n-m}(\mu_{12}(\theta)) \ ,
\nonumber
\eea
where a reduced mass is defined as 
\be
\mu_{12}(\theta) \equiv \sqrt{\mu_1^2+\mu_2^2+2\mu_1\mu_2\cos\theta} \ .
\label{mu12}
\ee
Performing the summation over $\nu$ and calculating the 
integral over $t$, the two flavor partition function 
for $\beta=1$ becomes 
\be
{\cal Z}^{(1)}(\theta,\mu_1,\mu_2) =
8\ \frac{I_2(\mu_{12}(\theta))}{\mu_{12}(\theta)^2} \ .
\label{Z1m12}
\ee
Let us stress the striking similarity to $\beta=2$ where 
\cite{Poul:1999,Lenaghan:2001ur} 
\be
{\cal Z}^{(2)}(\theta,\mu_1,\mu_2)=
2\ \frac{I_1(\mu_{12}(\theta))}{\mu_{12}(\theta)} \ .
\label{Z2m12}
\ee
At $\theta=\pi$, note that $\mu_{12}(\theta=\pi) = |\mu_1-\mu_2|$.  
To take the limit of degenerate quark masses, $\mu_1=\mu_2=\mu$, we simply
replace $\mu_{12}(\theta)=2\mu|\cos\frac{\theta}{2}|$ in
eqn.~(\ref{Z1m12}). This agrees with the results from 
Ref.\ \cite{Smilga:1995tb}. %and summing over all winding numbers.  
At $\theta=\pi$, we find ${\cal Z}^{(1)}(\theta=\pi,\mu,\mu) = 1$,
i.e. the partition function is independent of quark masses and volume
just like the partition function for $\beta=2$
\cite{Smilga,Lenaghan:2001ur}. In Ref.\ \cite{Smilga}, this was shown to be
due to a cancelation of terms at lowest order in chiral perturbation theory
and higher--order terms were considered.  In the Leutwyler--Smilga 
scaling regime, however, these terms are suppressed.

%%%%%%%%%%%%%%%%%%%%%%%%%%%%%%%%%%%%%%%%%%%%%%%%%%%%%%%%%%%%%%%%%%%%

%\subsection{${\fett \beta=4}$}
%\label{sec:Z4}

As discussed in Ref.\ \cite{Leutwyler:1992yt}, the allowed topological 
charges in adjoint QCD, $\beta=4$, are rescaled by a factor of $N_c$, so $\nu =
\bar\nu/N_c$ where $\bar{\nu}$ is an integer.  The partition function
thus only depends on the combination $M e^{i\theta/({\bar N_f}N_c)}$.
The number of Majorana fermions is given by ${\bar N_f}$ and 
we restrict the present discussion to $\bar N_f=2$.
%As above, we
%only present results for $\bar N_f=2$ here and refer to the appendix for
%larger numbers of flavors. 
The partition function with fixed topological charge can be calculated 
following \cite{Smilga:1995tb}:
%along the same lines as in \cite{Smilga:1995tb} leading to
\bea 
&&{\cal Z}_\nu^{(4)}(\mu_1,\mu_2) 
= \int_0^{2\pi} \!\!\frac{db}{4\pi} \!\sum_{n=-\infty}^{+\infty} 
\frac{1}{2n+1} 
\label{Z4m12}\\
&&\times\Big\{ I_{\bar{\nu}+n}(\mu(b)) I_{\bar{\nu}-n}(\eta(b)) 
             +I_{\bar{\nu}-n}(\mu(b)) I_{\bar{\nu}+n}(\eta(b)) \Big\} \ ,
\nonumber
\eea
with 
\bea
\mu(b) & =& \mu_1 \cos(b)^2 +\mu_2 \sin(b)^2\ , \nonumber \\
\eta(b)&= & \mu_2 \cos(b)^2 +\mu_1 \sin(b)^2 \ .
\label{meb}
\eea
Although the full partition function can be obtained 
from (\ref{Z4m12}) (see eqn.~(\ref{Z4m12full})), we only treat the degenerate mass case 
for simplicity in what follows. 
The partition function (\ref{Z4m12}) then drastically simplifies to 
\be
{\cal Z}_\nu^{(4)}(\mu,\mu)=\int_0^1\!dt I_{2\bar{\nu}}(2t\mu) \ .
\label{Z4nu}
\ee
This can be seen by applying $\partial_\mu \mu$ to
eqn.~(\ref{Z4m12}) at
equal mass $\mu$ %from \cite{Smilga:1995tb} 
and then integrating back.
In order to compute the summation over the topological charges $\bar{\nu}$, 
%winding numbers for the $\beta=$ 1 and 4 partition functions, 
it is useful to 
split the generating function for $I_{\nu}(x)$ into 
odd and even parts:
\be
\sum_{\nu=-\infty}^{\infty} t^{2\nu} I_{2\nu}(x) = 
        \cosh\left[\frac{x}{2}\left(t+t^{-1}\right)\right] \ ,
\ee
and similarly for the odd contribution. 
With the help of this identity and after integrating over $t$, we obtain 
\be
\label{Z4tmm}
{\cal Z}^{(4)}(\theta,\mu,\mu) =
\frac{\sinh(2\mu\cos\frac{\theta}{2})}{2\mu\cos\frac{\theta}{2}} \,\, ,
\ee
which was first derived in Ref.\ \cite{Leutwyler:1992yt}.
At $\theta=\pi$, the partition function is again equal to unity and 
thus is independent of quark masses and volume.  This is again 
an indication that there are dominant terms at next--to--leading order in chiral 
perturbation theory. % which are suppressed in Leutwyler--Smilga  scaling regime.
For ${\bar N_f}>2$ we refer to the appendix.

%%%%%%%%%%%%%%%%%%%%%%%%%%%%%%%%%%%%%%%%%%%%%%%%%%%%%%%%%%%%%%%%%%%%%%%%%%
\section{Topological Density and Chiral Condensate}
\label{sec:dlogZ}

The chiral condensate and the topological density are defined as 
\bea 
\Sigma_j^{(\beta)}(\theta,\{\mu_i\}) &\equiv& \Sigma%\frac{1}{V} 
        \frac{\partial}{\partial \mu_j} 
        \log {\cal Z}^{(\beta)}(\theta,\{\mu_i\}) \,\,\, ,\\ 
\sigma^{(\beta)}(\theta,\{\mu_i\}) &\equiv& -\frac{1}{V} 
        \frac{\partial}{\partial \theta} 
        \log {\cal Z}^{(\beta)}(\theta,\{\mu_i\}) \ .
\eea
In the case where ${\cal Z}$ is restricted to be 
a function of only some combination of 
$\theta$ and $\{m_i\}$, e.g. $X(\theta,\mu_1,\mu_2)=\mu_{12}(\theta)$, 
we readily find that
\bea 
\frac{\Sigma_j^{(\beta)}(\theta,\{\mu_i\})}{\sigma^{(\beta)}
(\theta,\{\mu_i\})} 
&=&
-V\Sigma\ \frac{\d_{\mu_j} X }{\d_\theta X} \ .
\eea
While this relation is a simple consequence of the observation
that ${\cal Z}={\cal Z}(X)$, it establishes a deep relationship 
between the chiral and the topological properties of the 
vacuum.  
The chiral and the topological susceptibility are also related although 
they are no longer proportional.

The $N_f=1$ case is special 
since the entire chiral symmetry group 
is explicitly broken by the axial 
anomaly and there is no spontaneous breaking of any symmetry.
As a result, there are no Goldstone modes. The topological density is 
\be \label{eq:topdennf1}
\sigma^{(\beta)}(\theta,\mu) = \frac{\mu}{V} \sin\theta \,\, ,
\ee
for $\beta=1,2$ and 4 (with $\theta\to\theta/N_c$) and there is no
discontinuity at $\theta=\pi$. This is another indication of the
interplay between the SB$\chi$S and the topological properties of the
theory.

Turning to the more interesting case of two degenerate flavors we only
have to calculate partial derivatives of the variable
$X=2\mu|\cos\frac{\theta}{2}|$ and of the corresponding partition
function. For $\beta=1$ we obtain 
\be 
\Sigma^{(1)}(\theta,\mu,\mu) =
\Sigma\ \frac{X}{\mu}\ \frac{I_3(X)}{I_2(X)}
\label{S1m} \ , 
\ee
and thus
\be
\sigma^{(1)}(\theta,\mu,\mu) = \frac1V\ \frac{\mu^2\sin\theta}{X} 
\ \frac{I_3(X)}{I_2(X)}\ .
\label{s1m}
\ee
The similarity with $\beta=2$ in \cite{Lenaghan:2001ur} 
is again striking. To recover their result 
we merely have to replace the logarithmic derivative 
$\partial_X\log{\cal Z}(X)$ 
leading to $I_2(X)/I_1(X)$ instead of $I_3(X)/I_2(X)$.

For $\beta=4$ we obtain from eqn.~(\ref{Z4tmm}) 
\bea
\Sigma^{(4)}(\theta,\mu,\mu) &=& \Sigma\ \frac{X\cosh X-\sinh X}{\mu\sinh X}
\ ,\\
\sigma^{(4)}(\theta,\mu,\mu) &=& \frac1V\ \frac{\mu\sin\frac{\theta}{2}}{X}\ 
\frac{X\cosh X-\sinh X}{\sinh X}
\label{S4m} \ .
\eea
For all three values of $\beta$ the quantities 
$\Sigma^{(\beta)}(\theta,\mu,\mu)$ and 
$\sigma^{(\beta)}(\theta,\mu,\mu)$ vanish at $\theta=\pi$.  
Furthermore, 
$\sigma^{(\beta)}(\theta,\mu,\mu)$ exhibits a discontinuity 
$\sim$ sign$(\cos\frac{\theta}{2})$  
at $\theta=\pi$ for large scaling variable, $\mu\gg1$. 
This follows from $\lim_{X\to\infty}I_{\nu+1}(X)/I_\nu(X)=$ sign$(X)$ for 
$\beta=1,2$ and similarly from eq. (\ref{S4m}) for $\beta=4$ 
(for $\beta=2$ this was shown already in \cite{Lenaghan:2001ur}). 
Thus Dashen's phenomenon is generic for all SB$\chi$B patterns with $N_f=2$.

The result for non-degenerate flavors follows in a similar fashion with 
$X=\mu_{12}(\theta)$. For $\beta=1$ and 2, 
the chiral condensate obtained from eqs. (\ref{Z1m12}) and (\ref{Z2m12}) 
respectively is easily seen to vanish at $\theta=\pi$. The case $\beta=4$ is
slightly more involved and we refer to the 
appendix (see eqs. (\ref{Z4m12pi}) and (\ref{Z4m12pidt})). 
To see the phase transition in $\sigma^{(\beta)}(\theta,\mu_1,\mu_2)$
for $\beta=1,2$
one must take the limit $\mu_1,\mu_2\rightarrow \infty$ 
such that 
\be
\lim_{\mu_1,\mu_2\to\infty}
\frac{(\mu_1-\mu_2)^2}{\mu_1\mu_2}\to0\ .
\label{limmu}
\ee 
This leads to 
$\mu_{12}(\theta)\to 2\sqrt{\mu_1\mu_2}|\cos\frac{\theta}{2}|$ and thus 
Dashen's phenomenon occurs in the same way as for degenerate masses. 
For $\beta=2$ the bound (\ref{limmu})
is consistent with the one found in \cite{DiVecchia:1980}.  
We obtain a discontinuity in $\sigma^{(\beta)}$ at $\theta=\pi$ and the vanishing of 
the chiral condensate for both degenerate and non-degenerate
  masses. Hence, these properties are not particular to the degenerate mass case.
We expect that Dashen's phenomenon also happens for $N_f>2$ for all
three values of $\beta$. It has been shown in \cite{Lenaghan:2001ur} to occur for
$N_f=3$ degenerate flavors and for $\beta=4$ with four flavours
it follows from eq. (\ref{beta4Nf4}).

%%%%%%%%%%%%%%%%%%%%%%%%%%%%%%%%%%%%%%%%%%%%%%%%%%%%%%%%%%%%%%%%%%%%%%%%%%
\section{Zeros of the partition functions}
\label{sec:zeros}

We now consider the zeros of the various partition functions in the
Leutwyler--Smilga regime\footnote{In Ref.\ \cite{kim}, an analysis of the zeros
was performed for the $\chi$RMT partition function of 
$\beta=2$ at fixed topological charge.}. 
Using the theorems of Yang and Lee
\cite{YL52} we expose the mechanism leading to 
Dashen's phenomenon in the
Leutwyler--Smilga regime for all $\beta$. 
Yang and Lee proved that the analytic behavior of the partition
function, ${\cal Z}$, is determined completely by its zeros when the
real parameters of the theory are continued into the complex
plane. The non-analytic behavior of ${\cal Z}$ as function of the
real-parameters occurs where the zeros in the complex plane pinch the
real axis. 
The non-analyticity of the partition function is tantamount to
the existence of a phase transition.  The SB$\chi$S in QCD occurs when
the zeros in complex quark mass plane pinch the real $\mu$-axis at
$\mu=0$. Dashen's phenomenon manifests itself as the zeros in the complex
$\theta$-plane pinch the real $\theta$-axis.
We note that since there is no spontaneous breaking 
of chiral symmetry for $N_f=1$, there is no solution to the 
equation ${\cal Z}^{\beta,N_f=1}(\theta,\mu)=0$ as is easily verified
using (\ref{Nf=1}).

For illustrative purposes, we next take 
{\sl two degenerate flavors} and {\sl any value of} $\beta$ and so
${\cal Z}^{(\beta)}(\theta, \mu,\mu)=
{\cal Z}^{(\beta)} (2\mu \cos\frac{\theta}{2})$. Consider the analytic
continuation of  $X\equiv2\mu\cos\frac{\theta}{2}$ into the complex plane.
For all three $\beta$'s the equation ${\cal Z}^{(\beta)}(X_j)=0$
is solved by purely imaginary values of $X_j$. 
The solutions are the non-vanishing zeros of $I_2$, $I_1$, and $\sinh$ 
for $\beta=1,2,4$ respectively.
Given the purely imaginary zeros, $X_j$, in the complex $X$-plane, it
is simple to find the zeros, $\mu_j$, in the complex $\mu$-plane given a real
$\theta$ 
\be
\mu_j = \frac{X_j}{2\cos\frac{\theta}{2}} \ ,
\ee
and the zeros, $\theta_j\equiv a_j+i b_j$, in the complex $\theta$-plane 
given a real $\mu$
\be
a_j=\pi \ \  {\rm and} \ \ b_j = 2 \, {\rm arcsinh}
\left(\frac{i X_j}{2\mu}\right) \ .
\ee 
From this we observe that $\mu_j\to \infty$ for $\theta\to\pi$, 
indicating the independence of the partition function on 
$\mu$ at $\theta=\pi$. Furthermore we see that
$\theta_j=\pi+i b_j \to \pi$ for $\mu\to\infty$, indicating Dashen's
phenomenon. We stress that this is independent of $\beta$. If we consider
{\sl two non-degenerate flavors}, then for $\beta=1$ and 2 we have
${\cal Z}^{(\beta)}(\theta,\mu_1,\mu_2)= 
{\cal Z}^{(\beta)}(\mu_{12}(\theta))$. Repeating the above argument we see
that we have exactly the same zeros, but in this case 
$X\equiv\mu_{12}(\theta)$. 
Solving this 
for $\theta_j$ at fixed real $\mu_1,\mu_2>0$ one finds that 
\be
a_j=\pi \ \  {\rm and} \ \ b_j = 
{\rm arccosh}\left(\frac{\mu_1^2+\mu_2^2-X_j^2}{2\mu_1\mu_2}\right) \ .  
\ee 
We observe that $\theta_j$ moves into the vicinity of $\pi$ as 
$\mu_1,\mu_2\to\infty$ provided that the bound (\ref{limmu}) holds,  
in complete accordance with what we found in the previous section.

In general, the partition function of QCD is not known in a closed
analytic form and consequently the zeros are less well understood. If
the partition function depends only on $Me^{i\theta/N_f}$, then the
zeros in the complex $\mu$-plane and the complex $\theta$-plane are
related, e.g. the zeros in the complex $\mu$-plane are simply rotated
around the origin when varying $\theta$.  However, from this relation alone
the accumulation of zeros in the complex $\theta$-plane in the vicinity of
$\theta=\pi$ and thus Dashen's phenomenon do not follow.

%%%%%%%%%%%%%%%%%%%%%%%%%%%%%%%%%%%%%%%%%%%%%%%%%%%%%%%%%%%%
\section{Summary}

We have shown that Dashen's phenomenon is generic for all three
S$\chi$SB patterns, $\beta=1,2$ and 4. By summing the finite volume
gauge theory partition functions over all topological charges in the Leutwyler-Smilga
regime, we have been able to directly calculate the
$\theta$-dependence of the corresponding chiral condensate and
topological density. For two flavors, these quantities were shown to be
proportional thus establishing a link between chiral and topological
properties of the underlying theory. At $\theta=\pi$ the chiral
condensate vanishes for degenerate and
non-degenerate quark masses and for all three $\beta$.  The
topological density has a discontinuity for all three values of
$\beta$ signaling the first--order phase transition predicted by
Dashen.  We have illuminated the mechanism by which the phase
transition occurs through an inspection of the zeros of the different
partition functions.

\acknowledgements
We thank P.H.~Damgaard and Michel Tytgat
for useful conversations and correspondence. 
The work of G.A. 
was supported in part by EU TMR grant ERBFMRXCT97-0122 and by 
the European network EUROGRID.

%%%%%%%%%%%%%%%%%%%%%%%%%%%%%%%%%%%%%%%%%%%%%%%%%%%%%%%%%%%%%%%%%%%%%%
\appendix

\section*{General $N_f$
%$\beta=4$ and $\bar N_f=4$
}

We will outline here how to derive the full partition function with $N_f>2$ 
flavors for $\beta=1$. The case $\beta=4$ works out along the same lines and 
expressions for $\Sigma$ and $\sigma$ follow subsequently. 
For fixed topological charge, we have \cite{Nagao:2000qn}
\bea
{\cal Z}_\nu^{(1)}(\{\mu\})&=&c^{(1)}_{N_f}
\frac{\prod_{f=1}^{N_f}\mu_f^\nu}{\Delta(\{\mu^2\})}
\left\{
\begin{array}{cc}
\mbox{Pf}(f) & N_f\ \mbox{even}\\
\mbox{Pf}\left(\begin{array}{cc}
f    & r\\
-r^T & 0
\end{array}\right) & N_f\ \mbox{odd}\\
\end{array}
\right. ,\nonumber\\
%\mbox{where} \label{Z1Nf}\\
f^{ij} &=& \int_0^1\! dt \ t^2\frac{I_{\nu-1}(t\mu_i)}{\mu_i^{\nu-1}} 
\frac{I_{\nu}(t\mu_j)}{\mu_j^{\nu}} \ -\ (i\leftrightarrow j)\ \ ,
\nonumber\\
r^i   &=& \frac{I_\nu(\mu_i)}{\mu_i^\nu} \ .
\label{Z1Nf}
\eea
$c^{(1)}_{N_f}$ 
is a $\nu$-independent constant and $\Delta$ is the Vandermonde determinant.
In Ref. \cite{Lenaghan:2001ur} sums over topological charge of the following type have 
been evaluated explicitly:
\be
{\cal B}= \sum_{\nu=-\infty}^\infty e^{i\nu\theta}
I_{\nu+l_1}(\mu_1)\cdots I_{\nu+l_{N_f}}(\mu_{N_f}) \ . 
\label{B}
\ee
We start with even $N_f$.  When rewriting Pf $=\sqrt{\det}$ above we can
pull the prefactor $\prod_{f=1}^{N_f}\mu_f^\nu$ inside the square root
and multiply each $i$th row of $\det f^{ij}$ with $\mu_i^\nu$ and each
$j$th column with $\mu_j^\nu$. We obtain that the factors of $\mu^\nu$
in $f^{ij}$ disappear.  We
can thus apply eqn.~(\ref{B}) after writing out the Pfaffian since the
only $\nu$-dependence left is in the Bessel functions. The results is 
a Pfaffian over different integrals of a single Bessel
function.  For $N_f$ odd, we only multiply the first $N_f$ rows and
columns of the matrix of size $N_f+1$ inside the Pfaffian.  In this
way, the matrix elements $r^i$ in the last column and row 
get multiplied only once with $\mu_i^\nu$,
which is precisely what we need to cancel $\mu_i^{-\nu}$.
We then proceed as for even $N_f$.

For $\beta=4$ and $\bar{N}_f=2$ we give the full partition function following 
from eqn.~(\ref{Z4m12}) for the sake of completeness: 
\bea
&&{\cal Z}^{(4)}(\theta,\mu_1,\mu_2) =
\int_0^{2\pi} \frac{db}{4\pi} \int_0^\pi \frac{da}{2} \sin(a) \nonumber\\
&&
\times\left\{ 
\cosh[\mu_{\mu\eta}(\theta)(e^{ia}T+e^{-ia}T^{-1})/2]
+ (\mu(b) \leftrightarrow \eta(b)) 
\right\}\!,
\nonumber\\
&&\ \ \ \ \ \ \ \ \ \ \ \ \ \ \ \ \ T =
\left[ 
\frac{e^{i\theta}\mu(b) +\eta(b)}{\mu(b)+e^{i\theta}\eta(b)}
\right]^{\frac12}\ ,
\label{Z4m12full}
\eea
where $\mu_{\mu\eta}(\theta)$ is the reduced mass 
of  $\mu(b)$ and $\eta(b)$  as in the definition (\ref{mu12}) and we have 
first performed the sum over topological charge.

At $\theta=\pi$ the partition function (\ref{Z4m12full}) 
simplifies drastically and we obtain
\bea
{\cal Z}^{(4)}(\pi,\mu_1,\mu_2)  &=& 
\int_0^{2\pi} \frac{db}{4\pi} \int_0^\pi\! da  
\nonumber\\
&\times&\sin(a) \cosh[\sin(a) (\mu_1-\mu_2)\cos(2b)] 
\nonumber\\
&=&\frac{\sinh(\mu_1-\mu_2)}{\mu_1-\mu_2} \ .
\label{Z4m12pi}
\eea
We can now easily see that the full chiral condensate 
$\Sigma^{(4)}(\theta)=\frac12\left(\Sigma^{(4)}_1(\theta)+\Sigma^{(4)}_2(\theta)\right)$ 
vanishes as it is obtained from 
$(\partial_{\mu_1} +\partial_{\mu_2} )\log {\cal Z}^{(4)}(\pi,\mu_1,\mu_2)$.
This is obviously zero for ${\cal Z}^{(4)}(\pi,\mu_1,\mu_2)$ 
being a function of $(\mu_1-\mu_2)$ only at $\theta=\pi$. 
The vanishing of $\sigma^{(4)}(\theta)$ can be seen from a similar 
calculation leading to 
\bea
0&=&
\left.\frac{\partial}{\partial\theta}
{\cal Z}^{(4)}(\theta,\mu_1,\mu_2)\right|_{\theta=\pi} 
\ = \int_0^{2\pi} \frac{db}{4\pi} \int_0^\pi\! da \label{Z4m12pidt}
\\
&\times&\sin(a)\cos(a)(\mu_1+\mu_2) \sinh[\sin(a) (\mu_1-\mu_2)\cos(2b)]\ ,
\nonumber 
\eea
which vanishes because of the first integral. 
For $\theta=0$ we get from (\ref{Z4m12full})
\be
{\cal Z}^{(4)}(0,\mu_1,\mu_2) =
\frac{\sinh(\mu_1+\mu_2)}{\mu_1+\mu_2} \ .
\ee
Along with (A4) and by analogy with $\beta=1$ and $\beta=2$ this suggests
that ${\cal Z}^{(4)}(\theta,\mu_1,\mu_2)=\sinh(\mu_{12}(\theta))/\mu_{12}(\theta)$
although we cannot prove this. If this is so then the analysis of section III
and IV for $\beta=1$ and $\beta=2$ with non-degenerate masses applies equally
well to $\beta=4$. 

Finally, we display the $\beta=4$ degenerate four-flavor case 
as an explicit example for ${\bar N_f}>2$:
\bea
&&{\cal Z}^{(4)}(\theta,\{\mu\}) = \frac{3}{\mu^2} \left(
-\frac{I_1(4\mu\cos\frac{\theta}{4})}{4\mu\cos\frac{\theta}{4}}
- \frac{I_1(4\mu\sin\frac{\theta}{4})}{4\mu\sin\frac{\theta}{4}}
\right.\label{beta4Nf4}
\\
&&\left.
+\int_0^1 \frac{dt}{2} 
\left[I_0\left(2\mu\sqrt{1+t^2+2t\cos\frac{\theta}{2}}\right)
                            + \left(\frac{\theta}{2}\to\frac{\theta}{2}+\pi\right)\right]
\right)\!. \nn
\eea
It can be most easily obtained using  
the $\chi$RMT correspondence \cite{Akemann:2000at}.

%%%%%%%%%%%%%%%%%%%%%%%%%%%%%%%%%%%%%%%%%%%%%%%%%%%%%%%%%%%%%%%%%%%%%%%%

\end{document}